\documentclass[twocolumn,showpacs,aps,prl,letter,footinbib,bibnotes,
superscriptaddress,groupedaddress]{revtex4}

\usepackage{amsmath,amssymb,amsbsy,graphicx,times,subfigure}
\usepackage{hyperref}
\usepackage[latin1]{inputenc}

\newtheorem{thm}{Theorem}

\newtheorem{lem}[thm]{\rm\em Lemma}

\newcommand{\id}{\mathbb{I}}
\newcommand{\ro}{\hat {\rho}}
\newcommand{\be}{\begin{equation}}
\newcommand{\ee}{\end{equation}}
\newcommand{\bea}{\begin{eqnarray}}
\newcommand{\eea}{\end{eqnarray}}

\newcommand{\tr}{{\rm Tr}\,}
\renewcommand{\det}{{\rm Det}\,}

\newcommand{\ket}[1]{|#1\rangle}
\newcommand{\bra}[1]{\langle#1|}

\newcommand{\eq}[1]{Eq.~(\ref{#1})}

\begin{document}

\title{Operational quantification of continuous variable correlations}
\author{Carles Rod\'o}
\affiliation{Grup de F\'isica Te\`orica, Universitat Aut\`onoma de Barcelona, 08193 Bellaterra (Barcelona), Spain.}
\author{Gerardo Adesso}
\affiliation{Grup de F\'isica Te\`orica, Universitat Aut\`onoma de Barcelona, 08193 Bellaterra (Barcelona), Spain.}
\affiliation{Dipartimento di Fisica ``E. R. Canianello'', Universit\'a degli Studi di Salerno, Via S. Allende, 84081 Baronissi (SA), Italy.}
\author{Anna Sanpera}
\address{Grup de F\'isica Te\`orica, Universitat Aut\`onoma de Barcelona, 08193 Bellaterra (Barcelona), Spain.}
\address{ICREA, Instituci\'o Catalana de Recerca i Estudis Avan\c cats, Barcelona, 08021 Spain.}
\date{November 12, 2007}

\begin{abstract}

We quantify correlations (quantum and/or classical) between two
continuous variable modes in terms of how many correlated bits can
be extracted by measuring the sign of two  local quadratures. On
Gaussian states, such `bit quadrature correlations' majorize
entanglement, reducing to an entanglement monotone for pure states.
For non-Gaussian states, such as photonic Bell states, ideal and
real de-Gaussified photon-subtracted states, and mixtures of pure
Gaussian states,
  the bit correlations
are shown to be a {\em monotonic} function of the negativity. This
yields a feasible, operational way to quantitatively measure
non-Gaussian entanglement in current experiments by means of direct
homodyne detection, without a full tomographical reconstruction of
the Wigner function.
\end{abstract}

\maketitle

Quantum information with continuous variables (CVs), relying on
quadrature entanglement as a resource, has witnessed rapid and
exciting progresses recently, also thanks to the high degree of
experimental control achievable in the context of quantum optics
\cite{books}. While Gaussian states (coherent, squeezed, and thermal
states) have been originally the preferred resources for both
theoretical and practical implementations, a new frontier emerges
with non-Gaussian states (Fock states, Schr\"odinger's cats, ...).
The latter can be highly nonclassical, possess in general more
entanglement than Gaussian states \cite{Wolf06}, and are useful to
overcome some limitations of the Gaussian framework such as
entanglement distillation \cite{NOGO} and universal quantum
computation \cite{Menic06}. Therefore,  it is of central relevance
to provide proper ways to quantify non-Gaussian entanglement in a
way which is experimentally accessible.

At a  fundamental level, the difficulty in the investigation of
entanglement -- quantum correlations -- can be traced back to the
subtle task of distinguishing it from classical correlations
\cite{groisman}. Correlations can be regarded as classical if they
can be induced onto the subsystems solely by local operations and
classical communication, necessarily resulting in a mixed state.  On
the other hand, if a pure quantum state displays correlations
between the subsystems, they are of genuinely quantum nature
(entanglement). We adopt here a pragmatic approach: if two systems
are {\em in toto} correlated, then this correlation has to be
retrieved between the outcomes of some local measurements performed
on them. We, therefore, investigate {\em quadrature correlations} in
CV states. We are also motivated by the experimental adequacy: field
quadratures can be efficiently measured by homodyne detection,
without the need for complete state tomography. Specifically, in
this paper, we study optimal correlations in bit strings obtained by
digitalizing the outcomes of joint quadrature measurements on a
two-mode CV system. First, we apply our procedure to Gaussian states
(GS), finding that
bit quadrature correlations provide a clear-cut quantification of
the total correlations between the two modes. They are monotonic
with the entanglement on pure states, and can be arbitrarily large
on mixed states, the latter possibly containing arbitrarily strong
additional classical correlations. We then address non-Gaussian
states (NGS), for which the exact detection of entanglement
generally involves measurements of  high-order moments
\cite{Shchukin05}.
The underlying idea is that for NGS obtained by de-Gaussifying an
initial pure GS and/or by mixing it with
a totally uncorrelated state, our measure {\em
based entirely on second moments} is still expected to be a
(quantitative) witness of the {\em quantum} part of correlations
only, {\it i.e.}~entanglement. We show that this is indeed the case
for relevant NGS including photon-subtracted states, photonic Bell
states, and mixtures of Gaussian states. Notably, the complete
entanglement picture in a recently demonstrated coherent
single-photon-subtracted state \cite{Ourjoumtsev07} is precisely
reproduced here in terms of quadrature correlations only. Our
results render non-Gaussian entanglement significantly more
accessible in a direct, practical fashion.

 \noindent {\bfseries \em  Quadrature measurements and bit
correlations.}--- We consider a bipartite CV system of two bosonic
modes, $A$ and $B$, described by an infinite-dimensional Hilbert
space.
The probability distribution associated to the
measurement of the rotated position quadrature  $\hat X_A(\theta)$
in mode $A$ with outcome $x_A^{\theta}$ and uncertainty $\sigma$ is
given by ${\cal P}_A(x_{A}^{\theta}) = \tr[\hat \rho_A \hat
R_A(\theta) \hat \sigma(x_{A}) \hat{R}_A(\theta)^{\dagger}] =
\tr[{\hat{R}_A(\theta)^{\dagger}} \hat \rho_A \hat{R}_A(\theta)
\hat{\sigma}(x_{A})]$, where $\hat \sigma(x_{A})$ is a single-mode
Gaussian (squeezed) state with first moments $\{ x_{A},0 \}$ and
covariance matrix ${\rm diag}\{\sigma^2,1/\sigma^2\}$.
Here $\hat R_A(\theta)$ is a unitary operator describing a rotation
of $\theta$ on mode $A$, corresponding to a symplectic
transformation given by
$R_A(\theta)={{\cos\theta\,\,-\sin\theta}\choose{\sin\theta\,\,\,\,
\cos\theta}}$ \cite{notenota}. Thus, one can either measure the
rotated quadrature on the state (passive view) or antirotate the
state and measure the unrotated quadrature (active view). Similarly
we define ${\cal P}_B(x_{B}^{\varphi})$ for mode $B$. The
probability distribution associated to a joint measurement of the
rotated quadratures $\hat{X}_A(\theta)$ and $\hat{X}_B(\varphi)$, is
given by ${\cal P}_{AB}(x_{A}^{\theta}, x_{B}^{\varphi}) = \tr[\hat
\rho_{AB} (\hat R_A(\theta) \otimes \hat R_B(\varphi)) (\hat
\sigma(x_{A}) \otimes \hat \sigma(x_{B})) ({\hat{R}_A(\theta)^
{\dagger}}\otimes{\hat{R}_B(\varphi)^{\dagger}})]$.

We digitalize the obtained outputs by assigning the bits $+\ (-)$ to
the positive (negative) values of the measured quadrature. This
digitalization transforms each joint quadrature measurement into a
pair of classical bits. A string of such correlated bits can be used
{\it e.g.}~to distill a quantum key \cite{Navascues05,Rodo07}. Let
us adopt a compact notation by denoting (at given angles
$\theta,\,\varphi$) ${\cal P}^{\pm}_{A} \equiv {\cal
P}_A(\pm|x_{A}^{\theta}|)$, and ${\cal P}^{\pm \mp}_{AB} \equiv
{\cal P}_{AB}(\pm|x_{A}^{\theta}|,\mp|x_{B}^{\varphi}|)$. The conditional
probability that
the bits of the corresponding two modes coincide
is given by ${P}^{=}_{AB} \equiv
({\cal P} ^{++}_{AB}+{\cal
P}^{--}_{AB})/\sum_{\{\alpha=\pm,\beta=\pm\}}{\cal
P}^{\alpha\beta}_{AB}$. Correspondingly, the conditional probability that they
differ is ${P}^{\neq}_{AB} \equiv ({\cal P}^{+-}_{AB}+{\cal
P}^{-+}_{AB})/\sum_{\{\alpha=\pm,\beta=\pm\}}{\cal
P}^{\alpha\beta}_{AB}$. Trivially, ${P}^{=}_{AB} + {P}^{\neq}_{AB} =
1$. If ${P}^{=}_{AB} > {P}^{\neq}_{AB}$ the measurement outcomes
display correlations, otherwise they display anticorrelations.
Notice that, if the two modes are completely uncorrelated,
${P}^{=}_{AB} = {P}^{\neq}_{AB} = 1/2$. For convenience, we
normalize the {\em strength} of bit correlations as
\begin{equation}\label{strength}
{\cal B}(|x_{A}^{\theta}|,|x_{B}^{\varphi}|) = 2 |{P}^{=}_{AB} -
1/2| = |{P}^{=}_{AB} - {P}^{\neq}_{AB}|\,,
\end{equation}
so that for a completely uncorrelated state ${\cal B}(|x_{A}^
{\theta}|,|x_{B}^{\varphi}|)=0$. The interpretation of \eq{strength}
in terms of correlations is meaningful if a {\em fairness} condition
is satisfied: on each single mode, the marginal probabilities
 associated to the outcomes ``$+$'' or ``$-$'' must be
the same: ${\cal P}^{+}_{A} = {\cal P}^{-}_{A}$, ${\cal P}^{+}_{B} =
{\cal P}^{-}_{B}$ \cite{notefair}.
%

For a two-mode CV system, whose state is described  by a Wigner
function $W$, we define the {\em `bit quadrature correlations'} $Q$
 as the average probability of obtaining a pair of
classically correlated bits (in the limit of zero uncertainty)
{\em optimized} over all possible choices of local quadratures
\begin{equation}\label{bitcorr}
Q(\ro)=\sup_{\theta,\varphi} {\int\!\!\int d x_A^\theta d
x_B^\varphi W(x_A^{\theta}, x_B^{\varphi}) [\lim_{\sigma \rightarrow
0}{\cal B}^\sigma(|x_{A}^{\theta}|,|x_{B}^{\varphi}|)]},
\end{equation}
where $W(x_A^{\theta}, x_B^\varphi) = \int\!\!\int d p_A^\theta d
p_B^\varphi W(x_A^\theta, p_A^\theta, x_B^\varphi, p_B^\varphi)$ is
the marginal Wigner distribution of the (rotated)  positions, and
$\{x_A^\theta, p_A^\theta, x_B^\varphi, p_B^\varphi\} = (R(\theta)
\oplus R(\varphi)) \{x_A, p_A, x_B, p_B\}$. After some algebra, we
can rewrite \eq{bitcorr} as $Q(\ro) = \sup_{\theta,\varphi}
|E_{A,B}^{\theta,\varphi}(\ro)|$, where
$E_{A,B}^{\theta,\varphi}(\ro)=\int\!\!\int d x_A^\theta d
x_B^\varphi {\rm sgn}(x_A^{\theta} x_B^{\varphi}) W(x_A^{\theta},
x_B^{\varphi})$ is the `sign-binned' quadrature correlation
function, which has been employed {\it e.g.}~in proposed tests of
Bell inequalities violation for CV systems \cite{notebell}. While
 this form is more suitable for an analytic
evaluation on specific examples, the
definition \eq{bitcorr} is useful to prove the following general properties of $Q$:\\
\begin{lem}[Normalization]\label{lemnorm}
\vspace*{-.5cm}\hspace*{-.1cm}: $0\le Q(\ro) \le 1$.\\
\em \noindent{\em Proof.\,} It follows from the definition of
$Q(\ro)$, as both ${\cal B}$ and the marginal Wigner distribution
range between $0$ and $1$. \hfill $\Box$
\end{lem}
\begin{lem}[Zero on product states]\label{lemprod}
\vspace*{-.2cm}\hspace*{-.1cm}: $Q(\ro_A \otimes \ro_B)=0$.\\
\em \noindent{\em Proof.\,} For a product state the probabilities
factorize {\it i.e.}~${\cal P}_{AB}^{\alpha \beta} = {\cal
P}_A^\alpha {\cal P}_B^\beta$ and so ${P}_{AB}^{=} =
{P}_{AB}^{\neq}$, where we have used the fairness condition. Namely
${\cal B} = 0$, hence the integral in \eq{bitcorr} trivially
vanishes. \hfill $\Box$
\end{lem}
\begin{lem}[Local symplectic invariance]\label{lemloc}
\vspace*{-.2cm}\hspace*{-.1cm}: Let  $\hat U_{A,B}$ be a unitary
operator amounting to a single-mode symplectic operation $S_{A,B}$
on the local phase space of mode $A,B$ {\em \cite{notenota}}. Then
$Q[(\hat U_A \otimes \hat U_B) \ro (\hat U_A^{\dagger} \otimes \hat
U_B^{\dagger})]=Q(\ro)$.\\
\em \noindent{\em Proof.\,} Any single-mode symplectic operation $S$
can be decomposed in terms of local rotations and local squeezings
(Euler decomposition). By definition \eq{bitcorr} is invariant under
local rotations, so we need to show that local squeezings, described
by symplectic matrices of the form $Z_r={\rm diag}\{1/r,r\}$, also
leave $Q(\ro)$ invariant.
Adopting the passive view, the action of
local squeezings  on the covariance matrix of each Gaussian state
$\hat \sigma(x^{A,B})$ is irrelevant, as we take eventually the
limit $\sigma \rightarrow 0$. The first moments are transformed as
$d_{A,B} \mapsto Z^{-1}_{s,t} d_{A,B}$, so that ${{\cal
B}}_{AB}^{\sigma\rightarrow 0}(|x_{A}|,|x_{B}|) \mapsto {{\cal
B}}_{AB}^{\sigma\rightarrow 0}(|s x_{A}|,|t x_{B}|)$. On the other
hand, the Wigner distribution is transformed  as $W(\xi) \mapsto
W[(Z^{-1}_s \oplus Z^{-1}_t) \xi)]$. Summing up, local squeezings
transform $\xi=\{x_A,p_A,x_B,p_B\}$ into $\xi_{st}=\{s x_A, p_A/s, t
x_B, p_B/t\}$. As  \eq{bitcorr} involves integration over the four
phase space variables $d^4\xi$, we change variables noting that $d^4
\xi = d^4 \xi_{st}$, to conclude the proof. $\hfill \Box$
\end{lem}
%

It follows from Lemmas (1--2) that if $Q > 0$, then the state
necessarily possesses correlations between the two modes. Lemma 3,
moreover, suggests that $Q$ embodies not only a qualitative
criterion, but might be interpreted as  {\em bona fide} operational
quantifier of CV correlations. We will now show that this is the
case for various important classes of states.
\begin{figure}[b!]
\includegraphics[width=5.4cm,height=2.3cm]{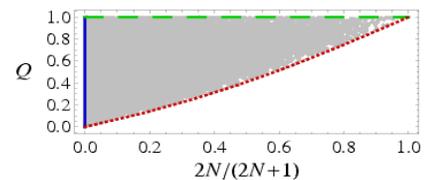}
\caption{\label{gaspacho} (color online) Bit quadrature correlations
vs normalized negativity for 18000 random two-mode Gaussian states.
The lowermost dotted (red) curve accommodates pure states. The
leftmost solid (blue) vertical line denotes separable states
parametrized by $c_p=0$, $c_x=\epsilon (\lambda^2-1)/\lambda$, with
$0\!\le\!\epsilon\!\le\!1$ and $\lambda_{a,b}=\lambda \rightarrow
\infty$. The uppermost dashed (green) horizontal line denotes
perfectly correlated states parametrized by
$c_x=(\lambda^2-1)/\lambda$, $c_p=\epsilon c_x$ (and $\epsilon$,
$\lambda$ as before). Product states (totally uncorrelated) lie at
the origin.}
\end{figure}

 \noindent {\bfseries \em
 Gaussian states.}---
Even though entanglement of GS is already efficiently accessible via
their covariance matrix \cite{books}, we
use such states as `test-beds' for understanding the role of
 $Q$
in discriminating CV correlations. The covariance matrix $\gamma$ of
any two-mode GS $\hat \rho$ can be written in standard form as:
$$ {\gamma}=
\left(\begin{array}{cc}
{\alpha}&{\delta}\\
{\delta}^{T}&{\beta} \end{array} \right), \quad \begin{array}{c}
\alpha=\lambda_a \id_2,\,\,\beta=\lambda_b \id_2, \\ \delta={\rm
diag}\{c_x,\,-c_p\}, \end{array}$$  where, without loss of
generality, we adopt the convention $c_x \ge |c_p|$.
The covariance matrix $\gamma$ describes a physical state if
$\lambda_{a,b}\ge 1$ and $\Delta \le 1+\det\gamma$, with
$\Delta=\det\alpha+\det\beta+2\det\gamma$. The {\em negativity}
\cite{Vidal02}, quantifying entanglement between the two modes,
reads $N(\hat \rho)=\max\{0,(1-{\tilde\nu})/(2{\tilde\nu})\}$, where
${\tilde\nu}^2 = [\tilde{\Delta}-(\tilde{\Delta}^2-4
\det\gamma)^{1/2}]/2$ with $\tilde{\Delta}=\Delta-4\det\delta$. For
two-mode GS, \eq{bitcorr} evaluates to:
\begin{equation}
\label{qpure} Q(\hat \rho) = (2/\pi) \arctan (c_x/\sqrt{\lambda_a
\lambda_b -c_x^2})\,,
\end{equation}
where the optimal quadratures are the standard unrotated positions
($\theta=\varphi=0$). First, we notice that $Q = 0 \Leftrightarrow
\rho$ describes a product state: for GS,  $Q > 0$ is then {\em
necessary and sufficient} for the presence of  correlations. Second,
we observe that for pure GS [reducible, up to local unitary
operations, to the two-mode squeezed states $\hat \rho_r \equiv
\ket{\phi_r}\!\bra{\phi_r}$ characterized by
$\lambda_{a,b}=\cosh(2r)$ and $c_x=c_p=\sinh(2r)$], \eq{qpure}
yields a monotonic  function of the negativity (see
Fig.~\ref{gaspacho}): $Q$ is thus, as expected, an operational {\em
entanglement measure} for pure two-mode GS. Third, we find that for
mixed states $Q$ {\em majorizes} entanglement. Given a mixed GS
$\hat \rho_N$ with negativity $N$, it is straightforward to see that
$Q(\hat \rho_N)$, [\eq{qpure}], is always greater than
$Q(\ket{\psi_N} \bra{\psi_N})=(2/\pi)
\arctan[({\tilde\nu}^2-1)/(2{\tilde\nu})]$, with $\ket{\psi_N}$
being a pure two-mode squeezed state with the same negativity $N$.
Hence  $Q$ quantifies  {\em total} correlations, and the difference
$Q(\hat\rho_N)-Q(\psi_N)$ (where the first term is due to total
correlations and the second to quantum ones) can be naturally
regarded as an operational measure of {\em classical} correlations
\cite{noteclassic}. We have evaluated $Q$ on random two-mode GS as a
function of their entanglement, conveniently scaled to $2N/(1+2N)$,
as shown
in Fig.~\ref{gaspacho}. Note that for any entanglement content there
exist maximally correlated GS  with $Q=1$, and also that  separable
mixed GS can achieve an arbitrary $Q$ from 0 to 1, their
correlations being only classical.
%
\begin{figure}[b!]{\centering{
\subfigure[]
{\includegraphics[width=4.3cm,height=2.5cm]{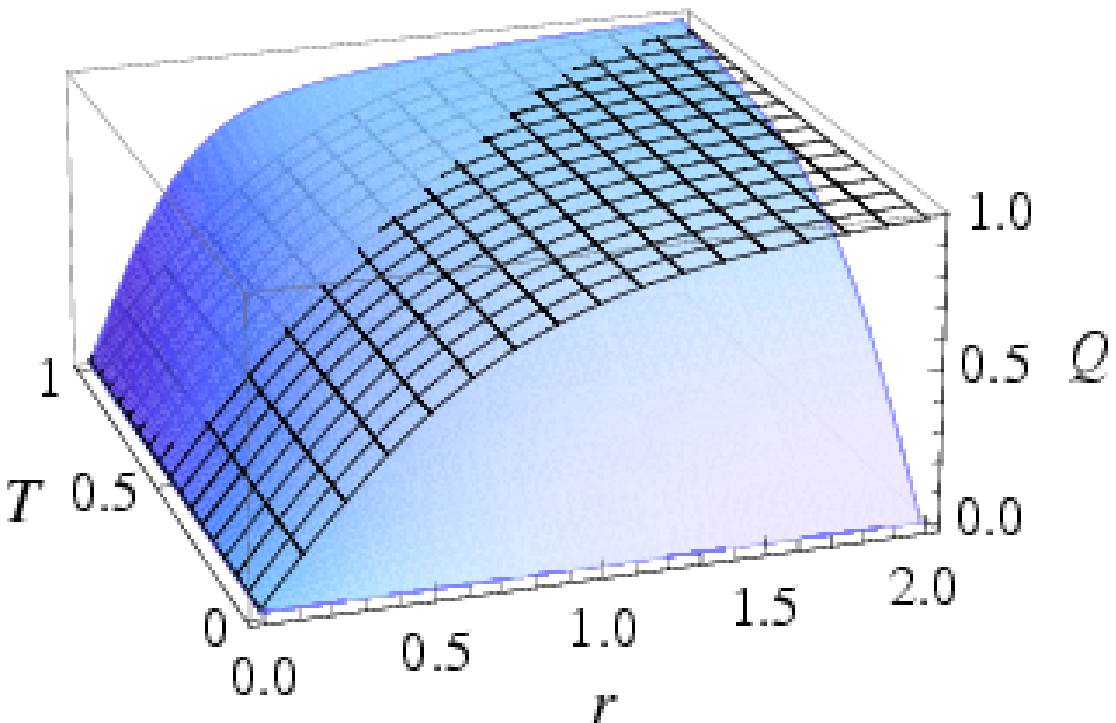}}
\hspace*{0.2cm} \subfigure[]
{\includegraphics[width=4cm,height=2.4cm]{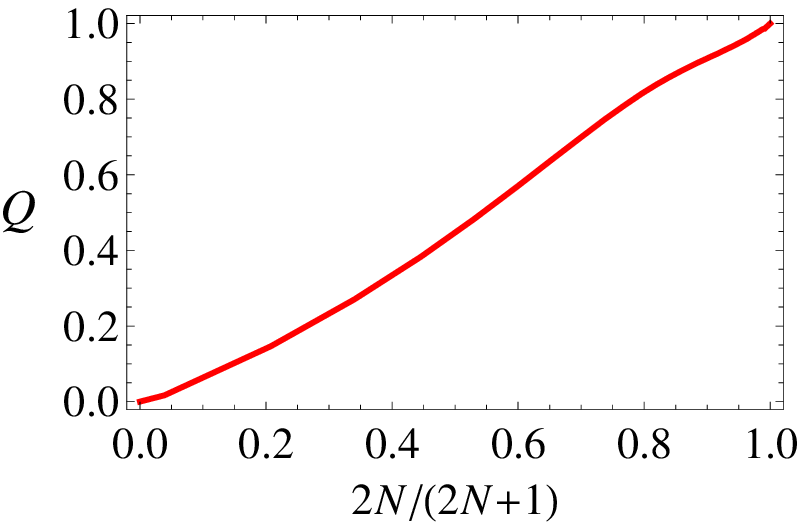}}
\caption{\label{fosof} (color online) (a) Bit quadrature
correlations $Q$ vs squeezing $r$ and beam splitter transmittivity
$T$ for photon-subtracted states (shaded surface) and two-mode
squeezed states (wireframe surface). (b) $Q$ vs normalized
negativity for photon-subtracted states.}}}
\end{figure}

 \noindent {\bfseries \em
 Non-Gaussian states.}---
Let us now turn our attention towards
NGS, whose entanglement and correlations are, in general, encoded in
higher moments as well. We focus on the most relevant NGS recently
discussed in the literature and sometimes experimentally realized.
Remarkably, we find for all of them a {\em monotonic} functional
dependence between the entanglement (negativity) and the quadrature
correlations $Q$.
%
\\ \noindent {\em Photonic Bell states.}
We consider Bell-like states of the form
$\ket{\Phi^{\pm}}=\sqrt{p}\ket{00}\pm \sqrt{1-p}\ket{11}$ and $\ket
{\Psi^{\pm}}=\sqrt{p}\ket{01}\pm\sqrt{1-p}\ket{10}$ (with $0\le
p\le1$), which are nontrivial examples of superpositions of Fock
states, entangled with respect to the (discrete) photon number.
The  negativity of these  states is $N_B=\sqrt{p(1-p)}$. For the
four of them, \eq{bitcorr} reads  $Q_B=(4/\pi) N_B$, showing a
perfect agreement between the quadrature CV correlations and the
entanglement.
\\ \noindent {\em Photon subtracted states.}
Most attention is being drawn by those CV states obtained from
Gaussian states via subtracting photons
\cite{Ourjoumtsev07,Furasawa}: they perform better as resources for
protocols like teleportation \cite{Kitagawa06,illu} and allow for
loophole-free tests of nonlocality \cite{bellref}. Let us recall
their preparation, following \cite{Kitagawa06}. The beam $A$ ($B$)
of a two-mode squeezed state $\ket{\phi_r}$ is let to interfere, via
a beam splitter of transmittivity $T$, with a vacuum mode $A^\prime$
($B^\prime$). The output is a four-mode Gaussian state of modes $A
A^\prime B B^\prime$. Detection of one photon in each of the two
beams $A^\prime$ and $B^\prime$, conditionally projects the state of
modes $AB$ into a pure NGS, given in the Fock basis by
$\ket{\psi_{ps}^{(T)}} = \sum^{\infty}_{n=0} c_n \ket{n,n}$, where
$c_n (\Lambda, T) = (n+1) (T \Lambda )^n \left(1-T^2
\Lambda^2\right)^{3/2}/\sqrt{T^2 \Lambda ^2+1}$, and $\Lambda =
\tanh r$. The limiting case $T\rightarrow 1$, occurring with
asymptotically vanishing probability, corresponds to an ideal
two-photon substraction, $\ket{\psi_{ps}^{(1)}}  \propto (\hat a _A
\hat a _B)\ket{\phi_r}$. For any $T$, the negativity is computable
as $N(\psi_{ps}^{(T)})=2/(1 - T \Lambda) - 1/(1 + T^2 \Lambda^2) -
1$. It increases with $r$ and with $T$: only in the case $T=1$ it
exceeds $N(\phi_r)$ for any $r$, diverging for $r\rightarrow
\infty$. For all $T<1$, the entanglement of $\ket{\psi_{ps}^{(T)}}$
eventually saturates, and above a squeezing threshold $r_T \gg 1$,
the original GS is more entangled than the resulting non-Gaussian
one. Following \cite{bellref}, the explicit expression of the
quadrature bit correlations \eq{bitcorr} can be analytically
obtained:
$$Q(\psi_{ps}^{(T)})=  \sum_{n>m=0}^{\infty}{\frac{2^{m+n+3} \pi
[{\cal F}(m,n)-{\cal F}(n,m)]^2 c_m c_n}{(m-n)^2 m! n!}},$$ where
the $c_k$'s are defined above, ${\cal F} = \left[\Gamma
\left(-\frac{m}{2}\right) \Gamma
\left(\frac{1-n}{2}\right)\right]^{-1}$, and $\Gamma$ is the Euler
function.
 As depicted in  Fig.~\ref{fosof}, the behaviour of the
entanglement is fully reproduced by
$Q(\psi_{ps}^{(T)})$
which again is a monotonic function of $N$: \eq{bitcorr} is thus
measuring truly quantum correlations of this important class of NGS
as well.
\\ \noindent {\em Experimental de-Gaussified states.}
\begin{figure}[b!]
\includegraphics[width=6.2cm,height=3cm]{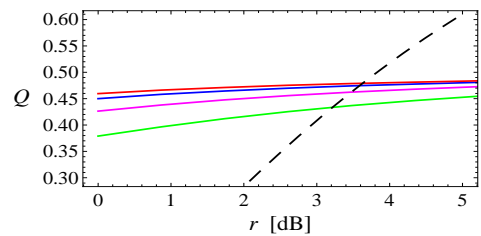}\\
\caption{\label{granapadano} (color online) Bit quadrature
correlations $Q$ vs squeezing $r$ (in decibels) for the
de-Gaussified states demonstrated in \cite{Ourjoumtsev07} with beam
splitter reflectivities $R$ equal to (from top to bottom): $3\%$
(red), $5\%$ (blue), $10\%$ (magenta), $20\%$ (green). The dashed
black curve depicts $Q$ for a two-mode squeezed state. Cfr.~Fig.~6
of \cite{Ourjoumtsev07}.}
\end{figure}
Up to now we considered nearly-ideal non-Gaussian situations. We
apply now our definition to the coherent photon-subtracted state
$\hat \rho_{exp}$ recently studied and demonstrated by Ourjoumtsev
{\it et al.} \cite{Ourjoumtsev07}. Referring to their paper for
details on the state preparation, we just mention that each of the
two beams of the two-mode squeezed state  $\ket{\phi_r}$ is let to
interfere with the vacuum at a beam splitter with reflectivity $R
\ll 1$, and by using an avalanche photodiode a single photon is
subtracted from the state  in a delocalized fashion. The realistic
description of the obtained highly mixed state involves many
parameters: we fix all  to the values obtained from the theoretical
calculations and/or experimental data of Ref.~\cite{Ourjoumtsev07},
but for the reflectivity $R$ and the initial squeezing $r$ which are
kept free. We then evaluate \eq{bitcorr} as a function of $r$ for
different values of $R$. Unlike the previous cases, optimal
correlations in the state $\hat \rho_{exp}$ occur between momentum
operators ($\theta=\varphi=\pi/2$). Also for this realistic mixed
case, the
  correlation measure  $Q$  reproduces precisely the
behaviour of the negativity, as obtained in \cite{Ourjoumtsev07}
after {\em full Wigner tomography} of the produced state $\hat
\rho_{exp}$. In particular, the negativity (and $Q$) increases with
the squeezing $r$, and decreases with $R$. Below a threshold
squeezing which ranges around $\sim$ 3 dB, the NGS exhibits more
entanglement (larger $Q$) than the original two-mode squeezed state.
Our results depicted in Fig.~\ref{granapadano} compare extremely
well to the experimental results (Fig.~6 of \cite{Ourjoumtsev07})
where the negativity is plotted as a function of $r$ for different
$R$'s.
%
%
\\ \noindent{\em Mixtures of Gaussian states etc.}
Recent papers \cite{VanLoock,mista}
dealt with mixed NGS of the form $\hat \rho_m = p \ket
{\phi_r}\!\bra{\phi_r} + (1-p) \ket{00}\!\bra{00}$, with $0 \le p
\le 1$. They have a positive Wigner function yet they are NGS (but
for the trivial instances $p=0,1$). Clearly, the de-Gaussification
here reduces entanglement and correlations in general.
The negativity of such states reads $N_m(\ro_m)=p N(\phi_r)= p({\rm
e}^{2r}-1)/2$ and is increasing both with $r$ and with $p$. The same
dependency holds for the bit correlations, $Q(\ro_m)=(2p/\pi)
\arctan\left\{N_m
   \left[\frac{1}{2
   N_m+p}+\frac{1}{p}\right]\right\}$, which again is a   monotonic
function of $N_m$ for any $p$. 
We further studied other NGS including photon-added and squeezed
Bell-like states \cite{illu}, and their mixtures with the vacuum:
for all we found a direct match between entanglement and $Q$
\cite{inprep}.

 \noindent {\bfseries \em Discussion.}---
 By analyzing the maximal number of correlated bits ($Q$) that
can be extracted from a CV state via quadrature measurements, we
have provided an operational quantification of the entanglement
content of several relevant NGS (including the useful
photon-subtracted states). Crucially, one can experimentally measure
$Q$ by direct homodyne detections (of the quadratures displaying
optimal correlations only), in contrast to the much more demanding
full tomographical state reconstruction. One can then easily invert
the (analytic or numeric) monotonic relation between $Q$ and the
negativity to achieve a {\em direct entanglement quantification}
from the measured data. Our analysis demonstrates the rather
surprising feature that entanglement in the considered NGS  can thus
be detected and experimentally quantified {\em with the same
complexity} as if dealing with GS.

Interestingly, this is not true for {\em all} CV states.   By
definition, $Q$ quantifies correlations encoded in the second
canonical moments only. We have realized that there exist also
states [{\it e.g.}~the photonic qutrit state
$\ket{\psi_h}=\ket{00}/\sqrt{2}+(\ket{02}+\ket{20})/2$] which,
though being totally uncorrelated up to the second moments ($Q=0$),
are strongly entangled, with correlations embedded only in higher
moments. The  characterization of such states is an intriguing topic
for further study \cite{inprep}. In this respect, it is even more
striking that the measure considered in this paper, based on (and
 accessible in terms of) second moments and homodyne
detections only, provides such an {\em exact quantification}  of
entanglement in a broad class of pure and mixed NGS, whose quantum
correlations are encoded nontrivially in higher moments too, and
currently represent the preferred resources in CV quantum
information. We focused on optical realizations of CV systems, but
our framework equally applies to collective spin components of
atomic ensembles \cite{polzik04}, and radial modes of trapped ions
\cite{referee}.
%
%
\\ \noindent We thank Ph.~Grangier, A.~Ourjoumtsev, and M. Piani for discussions.
We were supported by EU IP Programme SCALA, ESF PESC QUDEDIS, MEC
(Spanish government)  contract FIS2005-01369, CIRIT (Catalan
government) contract SGR-00185, and Consolider-Ingenio 2010
CSD2006-0019 QOIT.


\begin{thebibliography}{9927}
\bibitem{notenota}
Unitary operations $\hat U$ at most quadratic in the creation /
annihilation operators amount to symplectic transformations $S:S
\Omega S^T = \Omega$ in phase space, where the symplectic matrix
$\Omega$ encodes the canonical commutation relations. Then: $\hat
\rho \mapsto \hat U \hat \rho \hat U^{\dagger}$ (on the density
matrix) corresponds to $W(\xi) \mapsto W(S^{-1} \xi)$ (on the Wigner
function, where $\xi = \{x_1,p_1,\ldots,x_n,p_n\}$).
\bibitem{notefair}
Fair states have necessarily zero first moments, which can be
assumed without loss of generality. In general, one could take in
\eq{bitcorr} the difference between ${\cal B}$ computed on $\ro$ and
${\cal B}$ computed on $\ro_A\otimes\ro_B$. The latter is zero on
fair states.
\bibitem{notebell}
A homodyne Bell test  requires measuring two different rotated
quadratures per mode, to achieve violation of the bound
\cite{bellref}
$|E^{\theta,\phi}_{A,B}+E^{\theta',\phi}_{A,B}+E^{\theta,\phi'}_{A,B}-E^{\theta',\phi'}_{A,B}|\le
2$.
Here, we propose to measure a single quadrature per mode, which
displays one-shot optimal correlations, unveiling a powerful
quantitative connection with   Gaussian and non-Gaussian
entanglement.
\bibitem{noteclassic} The emerging measure of classical
correlations is special to Gaussian states, where the correlated
degrees of freedom are
 the field quadratures. Different approaches
to the quantification of classical vs quantum correlations were
proposed \cite{groisman}.
\end{thebibliography}
\end{document}